\documentclass[onecolumn,showpacs,preprintnumbers,amsmath,amssymb]{revtex4}
\usepackage{graphicx}
\usepackage{dcolumn}

\def\prl#1#2#3{{ Phys.   Rev.   Lett.  } {\bf #1}, #2 (#3)}

\def\pre#1#2#3{Phys.   Rev.   E {\bf #1}, #2 (#3)}

\def\jsp#1#2#3{J.   Stat.   Phys.   {\bf #1}, #2 (#3)}

\def\epl#1#2#3{Europhys. Lett. {\bf #1}, #2 (#3)}

\def\jsm#1#2#3#{J.Stat.Mech. {\bf#1}, #2 (#3)}

\def\noi{\noindent}
\def\bc{\begin{center}}
\def\ec{\end{center}}
 \newcommand{\bea}{\begin{equation}}
 \newcommand{\eea}{\end{equation}\noi}
 \newcommand{\ber}{\begin{eqnarray}}
 \newcommand{\eer}{\end{eqnarray}\noi}
\begin{document}
\title{Asymmetry of the work probability distribution}
\author{Arnab Saha}\email{www.arnab.bose.res.in}
\affiliation{S.N.Bose National Centre For Basic Sciences, JD-Block,
Sector-III, Salt Lake City, Kolkata-700098, India} 
\author{J.K. Bhattacharjee}\email{tpjkb@mahendra.iacs.res.in}
\affiliation{Department of Theoretical Physics,
Indian Association for the Cultivation of Science \\
Jadavpur, Calcutta 700 032, India}
\date{\today}
\begin{abstract}
We show, both analytically and numerically, that for a nonlinear system making a transition from one equilibrium state to another under the action of an external time dependent force, the work probability distribution is in general asymmetric.
\end{abstract}

\maketitle

We consider a system in contact with a heat bath, which is driven out of equilibrium by an external time dependent force. This force drives it from an equilibrium state A to another equilibrium state B. It was shown by Jarzynski  \cite{1,2,3} that the equilibrium free energy difference, $\Delta F$ between these states can be related to the probability distribution of the work done $W$ in taking the system from A to B. In particular, 
\bea
e^{\frac{-\Delta F}{KT}}=\left<e^{\frac{-W}{KT}}\right> 
\eea
where $K$ is Boltzmann constant and $T$ is the temperature. A related equality is that due to Crooks \cite{4,5} which relates the probabilities for the forward process $(A\rightarrow B)$ to the backward process $(B\rightarrow A)$. A couple of years ago a simple but effective experiment on a mechanical oscillator was done by Douarche, Ciliberto, Patrosyan and Rabbiosi (DCPR)\cite{6,7} who showed that for a Gaussian distribution of $W$, Eq (1) can be recast as 
\bea
\Delta F=\left<W\right >-\frac{\left<(W-\left<W\right>)^2\right >}{2KT}
\eea 
 
  This relation which is similar to a relation found by Landau and Lifshitz \cite{8} in the context of thermodynamic perturbation theory, was demonstrated experimentally and also analytically for a particular kind of forcing by DCPR for linear oscillators. The particularly convenient form of Eq (2) made us investigate whether it holds for nonlinear systems. In what follows, we have looked at the general system (although in highly viscous limit). Very recently for a quadratic $V(x)$, the system has been explored by Douarche et al\cite{9}.
\bea
m\ddot x+k\dot x=-\frac{\partial V}{\partial x}+M(t)+f(t)
\eea

where $M(t)$ is an externally applied time dependent force and $f(t)$ is a random force that allows the system to be in equilibrium in absence of $M(t)$. The system is supposed to be in equilibrium (state A) at $t=0$ and then we switch on $F(t)$ for a time $\tau$, after which $F(t)$ takes a constant value $F(\tau)$. The system reaches to state B and equilibriates. In going from state A to state B, the work done is 
\bea
W=-\int_{0}^{\tau}\dot M(t) x(t) dt
\eea
We are interested in the moments of W. We will work in the highly damped limit where the inertial term in Eq.(3) can be dropped. For a quadratic $V(x)$ (linear system), we will prove Eq.(2) for an arbitrary $M(t)$ and then go on to show that Eq.(2) needs to be modified for arbitrary $V(x)$. The most significant finding is that, even for a symmetric $V(x)$, the odd moments of $\Delta W=W-\left<W\right>$ are non-vanishing and hence the distribution of $\Delta W$ is asymmetric for all non-quadratic $V(x)$. We will show this analytically, using perturbation theory; suggest a generalization of Eq.(2)  and numerically establish that the probability distribution $P(\Delta W)$ is indeed asymmetric in general.
\\We begin with the linear harmonic oscillator under the action of a deterministic force $M(t)$ and random force $f(t)$. In the highly viscous limit, the system evolves according to 
\bea
\dot x+\Gamma x=M(t)+f(t)
\eea 
where the random force $f(t)$ has the correlation function 
\bea
\left<f(t)f(t^{\prime})\right>=2KT\delta(t-t^{\prime})
\eea
We calculate the moments of the work done, from above dynamics. The solution for $x(t)$ can be written down as 
\bea
x(t)=\int G(t-t^{\prime})[M(t^{\prime})+f(t^{\prime})]dt^{\prime}
\eea
where $G(t-t^{\prime})$ is the causal Green function,
\bea
G(t-t^{\prime})=\Theta(t-t^{\prime})e^{-\Gamma (t-t^{\prime})}
\eea
Clearly, the average of the work is 
\bea
\left<W\right>=-\int^{\tau}dt_1\dot M(t_1)\int^{t_1}G(t_1-t_2)M(t_2)dt_2
\eea
while, the deviation from the average is 
\bea
\Delta W=W-\left<W\right>=-\int^{\tau}dt_1\dot M(t_1)\int^{t_1}G(t_1-t_2)f(t_2)dt_2
\eea
which has the mean square value of
\ber
\left<(\Delta W)^2\right>=2KT\int^{\tau}dt_1\dot M(t_1)\int^{\tau}dt_2\dot M(t_2)\\ \nonumber\int^{t_2}dt^{{\prime}{\prime}}\int^{t_1}dt^{\prime}G(t_1-t^{\prime})G(t_2-t^{{\prime}{\prime}})\delta(t^{\prime}-t^{{\prime}{\prime}})
\eer
where Eq.(6) has been used. Noting the identity, derived from the time translational invariance
\bea
\int_0^{t_2}G(t_2-t^{{\prime}{\prime}})G(t_1-t^{{\prime}{\prime}})dt^{{\prime}{\prime}}=G(t_1-t_2)/2\Gamma
\eea
we arrived at 
\bea
\frac{\left<(\Delta W)^2\right>}{2KT}=\frac{1}{\Gamma}\int^\tau dt_1\dot M(t_1)\int^t_1 dt_2\dot M(t_2)G(t_1-t_2)
\eea
Using the above and Eq.(9)
\ber
\left<W\right>-\frac{\left<(\Delta W)^2\right>}{2KT}=
-\int^\tau dt_1\dot M(t_1)\int^t_1 dt_2G(t_1-t_2)[M(t_2)+\frac{\dot M(t_2)}{\Gamma}]
\eer
Integrating by parts the first term in the integral above and using $G(0)=0$ due to causality, we find 
\bea
\left<W\right>-\frac{\left<(\Delta W)^2\right>}{2KT}=-\frac{M^2}{2\Gamma}
\eea
The free energy change is precisely this amount, and that establishes Eq.(2) for an arbitrary forcing M(t).\\
 We now consider the inclusion of a non linear term in the motion, which becomes 
\bea
\dot x+\Gamma x+\lambda x^3=M(t)+f(t)
\eea
The question to ask is whether the equality in Eq.(2) still holds? To investigate this, we specialize to the case of $M(t)=M_0t$ as studied by DCPR and carry out a perturbative calculation to $O(\lambda)$.\\
We write 
\bea
x=x_0+\lambda x_1+\lambda^2 x_2+...
\eea
and substituting in Eq.(16) and equating the coefficients of equal powers of $\lambda$ on either side we get 
\ber
\dot x_0+\Gamma x_0 &=& M(t)+f(t)\\ \nonumber
\dot x_1+\Gamma x_1 &=&-x_0^3
\eer
and so on. We can now expand $W(\tau)$ according to Eq.(4) and (17),
\ber
W =W_0+\lambda W_1+...
\eer
where 
\ber
W_0&=&-\int^{\tau}\dot M(t) x_0(t)dt\\ \nonumber
W_1&=&-\int^{\tau}\dot M(t) x_1(t)dt
\eer
From Eq.(18) we get
\ber
x_0&=&\int_0^t G(t-t^{\prime})[M(t^{\prime})+f(t^\prime)]dt^{\prime}\\ \nonumber
x_1&=&-\int_0^t G(t-t^{\prime})x_0^3dt^{\prime}
\eer
We note that in the way we set it up,$G(t_1-t_2)$ is exactly the same $G$ that we had for the linear problem. We have already calculated $\left<W_0\right>$ (Eq.(9)) and now we concentrate on $\left<W_1\right>$. We write
\ber
\nonumber \left<W\right>&=&\int^{\tau}\dot M dt\int^{t}dt^{\prime}G(t-t^{\prime})\left[\int^{t^\prime}dt_1G(t^{\prime}-t_1)M(t_1)\right]^3 \\ 
&+& 3\int^{\tau}\dot M dt\int^t dt^\prime G(t-t^\prime)\int^{t^\prime}dt_1dt_2dt_3 \\ \nonumber
&M(t_1)&G(t^\prime-t_1)G(t^\prime-t_2)G(t^\prime-t_3)\left<f(t_2)f(t_3)\right>
\eer
The second term in the r.h.s vanishes once we use Eq.(12) and causality. We are left with 
\bea
\left<W\right>=\int^{\tau}\dot M dt\int^{t}dt^{\prime}G(t-t^{\prime})\left[\int^{t^\prime}dt_1G(t^{\prime}-t_1)M(t_1)\right]^3
\eea
We now use $M(t)=M_0t$ and carry out the the integration to arrive at 
\bea
\left<W\right>=\frac{M_0^4}{\Gamma^3}\left[\frac{\tau^4}{4\Gamma}-\frac{2\tau^3}{\Gamma^2}+\frac{15\tau^2}{2\Gamma^3}-\frac{16\tau}{\Gamma^4}\right] 
\eea
,keeping the leading order terms, i.e. terms which increase with $\tau$.
\\ Let us now turn to the calculation of the variance $\left<(W-\left<W\right>)^2\right>$. The perturbative calculation generates the following upto $0(\lambda)$
\ber
\left<(\Delta W)^2\right>=\left<(\Delta W_0)^2\right>+2\lambda \left<\Delta W_0\Delta W_1\right>
\eer
Where $\Delta W_0=W_0-\left<W_0\right>$ and $\Delta W_1=W_1-\left<W_1\right>$. We have already calculated the first term in the r.h.s. Now we will concentrate on the second term. In the $O(\lambda)$ correction in the variance, we have found that the disconnected parts (i.e. where the averaging is over $\Delta W$ and $\Delta W_1 $ separately.) do not contribute. Specializing to the case $M(t)=M_0t$, calculation leads to the $O(\lambda)$ term of the variance 
\bea
\eea
Thus, corrected upto $O(\lambda)$ we get
\bea
0(\lambda)\phantom x part\phantom x of \frac{\left<(\Delta W)^2\right>}{2KT}=-\frac{M_0^4}{\Gamma^3}\left[\frac{2\tau^3}{\Gamma^2}-\frac{27\tau^2}{2\gamma^3}+\frac{16\tau}{\Gamma^4}\right]
\eea
,keeping the leading order terms. Thus,
\bea
\left<W\right>-\frac{\left<(\Delta W)^2\right>}{2KT}=\left<W_0\right>-\frac{\left<(\Delta W_0)^2\right>}{2KT}+\frac{\lambda M_0^4\tau^4}{4\Gamma^4}-\frac{6\lambda M_0^4\tau^2}{\Gamma^6}
\eea
Hence following Eq.(2) and above, we see that (i) $\Delta F$ is no longer $\left<W\right>-\frac{\left<(\Delta W)^2\right>}{2KT}$, and (ii) for $M=M_0t$, the difference arises at $O(\tau ^4)$ and $O(\tau ^2)$. (i) implies a non-Gaussian distribution for $W$ which perhaps not so surprising but it is (ii) which has the interesting consequence. (ii) leads to the conclusion that correction to the result has to have an asymmetric part. This is because, if the correction to the Gaussian distribution is symmetric, then the first correction that would arise is the flatness factor $(\left<(\Delta W)^4\right>-3\left<(\Delta W)^2\right>^2)$, but a rather long and tedious calculation reveals that leading contribution to the term is at $O(\tau)$. The contribution at $O(\tau^2)$ can come only from $\left<(\Delta W)^3\right>$. Does dynamics contribute to $\left<(\Delta W)^3\right>$ ? We note that 
\bea
\left<(\Delta W)^3\right>=3\lambda \left<(\Delta W_0)^2(\Delta W_1)\right>
\eea
and a cursory inspection of the solution of the equation of motion, we find that $\left<(\Delta W)^3\right>$ is nonzero and the leading term is $O(\tau^2)$.
\\ This observation inspires us to start with a work probability distribution $P(W)$ given by 
\bea
P(W)\propto e^{ [-\frac{(W-W_0)^2}{2\sigma^2}-\frac{\mu_1(W-W_0)^3}{\sigma^3}-\frac{\mu_2(W-W_0)^4}{\sigma^4}]}
\eea
where $\mu_1$, $\mu_2$ and $\sigma$ are parameters. Now we work out the expectation value of $e^{\frac{-W}{KT}}$ with the above distribution keeping $\mu_1$, $\mu_2$ small,and thus we have found 
\ber
&&\left<\exp\left[{\frac{-W}{KT}}\right]\right>=e^{-[\frac {W_0}{KT}+\frac{\sigma^2}{2(KT)^2}]}\times\\ 
\nonumber&&\left[1+\frac{\mu_1\sigma}{KT}\left(3+\frac{\sigma^2}{(KT)^2}\right) -\mu_2\frac{\sigma^2}{(KT)^2}\left(6+\frac{\sigma^2}{(KT)^2}\right)\right]
\eer
,keeping only the linear order terms of $\mu_1$ and $\mu_2$. We now calculate different expectation values, like $\left<(W-W_0)\right>$, $\left<(W-W_0)^2\right>$, $\left<(W-W_0)^3\right>$ and $\left<(W-W_0)^4\right>$ and replacing $\mu_1$, $\mu_2$ and $\sigma$ in above by these expectation values, we get the following basic result [using Eq.(2)] 
\ber
\Delta F&=&\left<W\right>-\frac{\left(W-\left<W\right>\right)^2}{2KT}+\frac{\left(W-\left<W\right>\right)^3}{6(KT)^2}\\
\nonumber &+& \frac{1}{24}\left[\frac{3\left<(W-\left<W\right>)^2\right>^2-\left<(W-\left<W\right>)^4\right>}{(KT)^3}\right]
\eer
To test the relation numerically, we decided to work first with the quadratic nonlinearity in
equation of motion,
\bea
\dot x+\Gamma x+\lambda x^2=M(t)+f(t)
\eea
We will restrict ourselves to such values of $\lambda$ with $\Gamma=1$ (we take $\Gamma=1$ every where), so that trajectory does not run away. In this case, the dynamics does not generate the last term in Eq.(32) to the lowest order in $\lambda$. It is the cubic deviation which is the most significant and that would imply an asymmetric probability distribution for the work $W$. This is not unexpected since the potential for Eq.(33) is cubic and hence asymmetric. We have taken,
\ber
M(t)&=&0 \phantom x\phantom x\phantom x t<0 \\ \nonumber &=&M_0t/\tau \phantom x\phantom x\phantom x0\leq t<\tau/M_0\\ \nonumber &=&1\phantom x\phantom x \phantom x t\geq \tau/M_0
\eer
We allow the system to equilibriate at $t=0$ and at $t=\frac{\tau}{M_0}$ by running it with M=0 for an interval and with $M=1$ for a similarly long interval. In between we generate the values of $x$ at different points by the following,
\bea
x(t+\Delta t)=x(t)-(x(t)+\lambda x^2(t))\Delta t+M(t)\Delta t+\sqrt{2KT\Delta t}\eta(t)
\eea
where $\eta(t)$ is a random number between $0$ and $1$. We calculate all the quantities in the unit of $2KT$. We calculate a trajectory $[x(t)]_{\tau}$ starting from an initial value and evaluate the work according to Eq.(4). The ensemble is one of initial conditions $x(0)$ and we calculate $\left<W\right>$, $\left<(\Delta W)^2\right>$, $\left<(\Delta W)^3\right>$. We have also found the work distribution function $P(W)$. For $\lambda=0$ (i.e. the system with quadratic potential) and for $\lambda=20$ the distributions are shown in fig.(1,3) respectively. For nonzero $\lambda$ the distribution is asymmetric, which is expected for an asymmetric potential. The free energy difference between times $t=0$ and $t=\tau$ can be found as
\bea
\Delta F=\frac{1}{2}x^2+\frac{\lambda}{3}x^3
\eea
It does not match with the r.h.s of Eq.(2) for nonzero $\lambda$ instead we calculate $\left<(\Delta W)^3\right>$ and find that Eq.(36) gives a correct description.
\\ We repeated the numerics with quartic oscillator (i.e.$V(x)=\frac{1}{2}x^2+\frac{\lambda}{4}x^4$)and found that $\left<(\Delta W)^3\right>$ is certainly nonzero, indicating an asymmetric distribution. For small values of $\lambda$, the asymmetry is striking. For large values of $\lambda$ due to dominating  $\left<(\Delta W)^4\right>$ the distribution becomes sharply peaked, and the asymmetry is difficult to make out, although its existence is guaranteed by the nonzero value of $\left<(\Delta W)^3\right>$. The distribution for $\lambda=0.1$ and $\lambda=20$ is shown in fig.(2,4) respectively. In fig(5) comparison between the work distribution functions for the cases when $\lambda=0.1$ and $\lambda=0$ are shown by plotting together. The asymmetry is clear from it.  Recently Mai and Dhar \cite {10}have found an asymmetric distribution for $V(x)=ax^2+bx^3+cx^4$. Our contention is that the asymmetry exists even if $b=0$. Here we have to calculate $\Delta F$ by
\bea
\Delta F=\frac{1}{2}x^2+\frac{\lambda}{4}x^4
\eea
 We have calculated $\Delta F$ using Eq.(32) also, after calculating required moments from the work probability distribution obtained, and have found again, that it gives the correct description.
\\
We end by pointing out a possible application. We consider a ferromagnet or an Ising magnet near but above its critical point $T_c$. We can imagine being close to $T_c$, but sufficiently far away so that the mean field Landau model is valid. If we now switch on an time dependent magnetic field, then the dynamics of the mean magnetisation will be given by an equation of the form shown by Eq.(16). If we are in the region $T<T_c$, then the dynamics will be governed by Eq.(16) with an added quadratic nonlinearity- the kind considering Ref.10. It will be interesting to check the veracity of Eq.(32) in this case.

\begin{figure}[h]
\centering
\includegraphics[scale=0.2,angle=270]{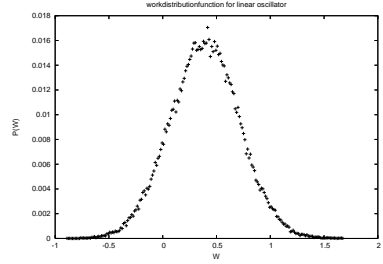}
\caption{Here force is applied for 8.4 sec. $M_0$ and $\tau$ are 100 and 840 respectively, in S.I. units. Here $\lambda=0$. Note the symmetry of the distribution}
\end{figure}

\begin{figure}[h]
\centering
\includegraphics[scale=0.2,angle=270]{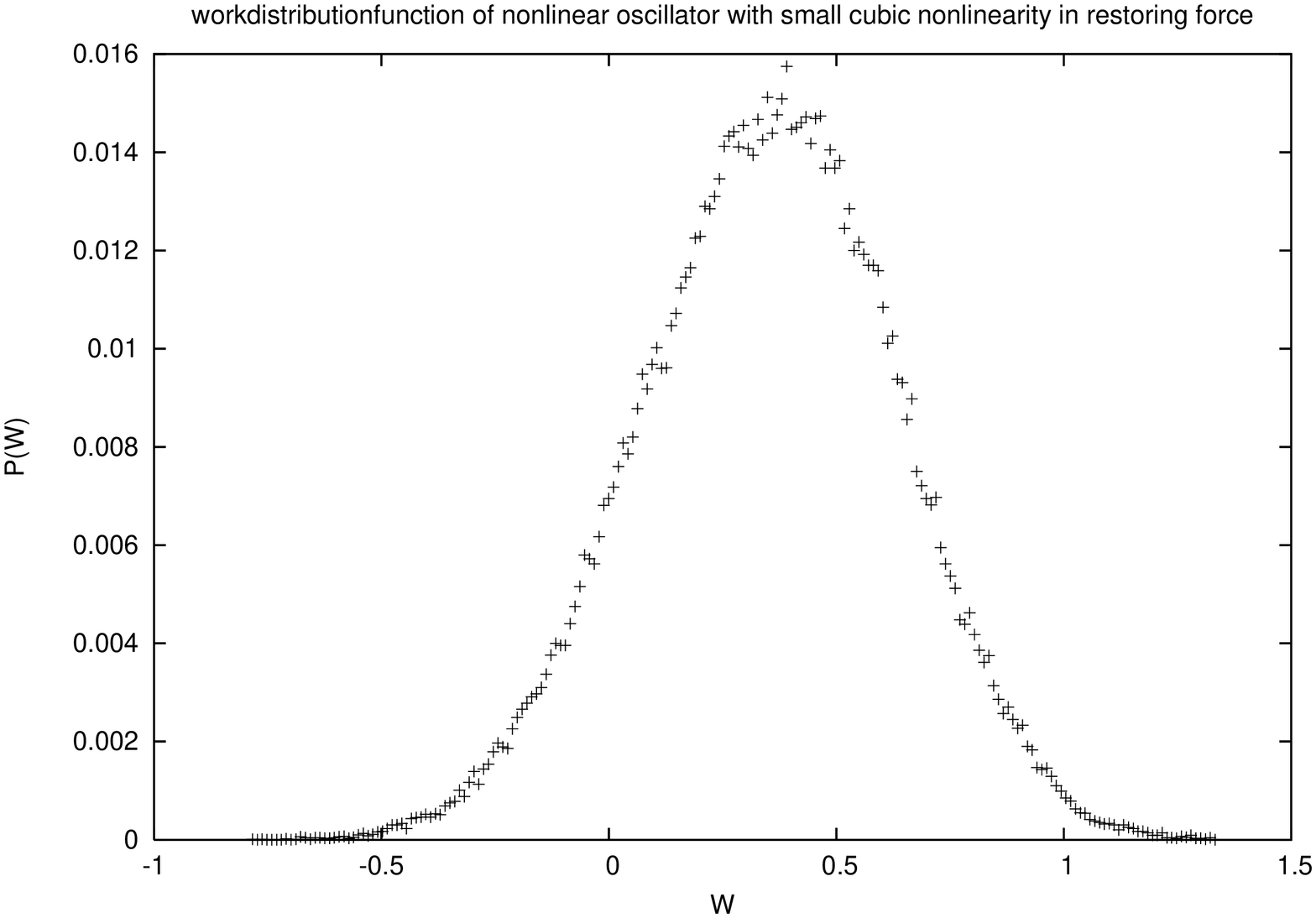}
\caption{Here force is applied for 8.4 sec. $M_0$ and $\tau$ are 100 and 840 respectively, in S.I. units. Here $\lambda=0.1$. The tail in left side signifies the asymmetrical nature of the distribution here. Here $V(x)=(1/2)x^2+(0.1/4)x^4$.}
\end{figure}


\begin{figure}[h]
\centering
\includegraphics[scale=0.2,angle=270]{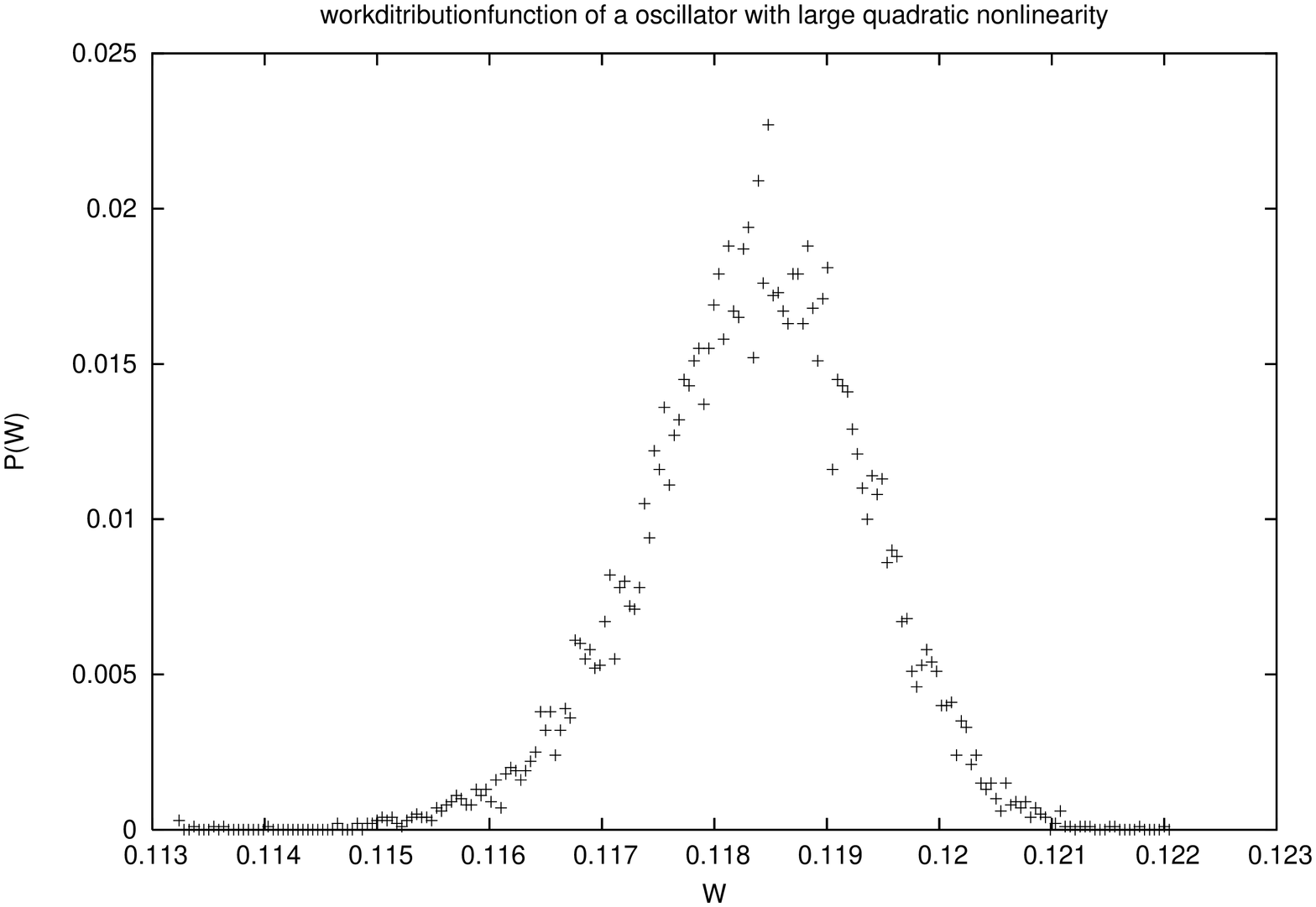}
\caption{Here force is applied for 8.4 sec. $M_0$ and $\tau$ are 100 and 840 respectively, in S.I. units. Here $\lambda=20$. The tail in left side signifies the asymmetrical nature of the distribution here. Here $V(x)=(1/2)x^2+(20/3)x^3$.}
\end{figure}


\begin{figure}[b]
\centering
\includegraphics[scale=0.2,angle=270]{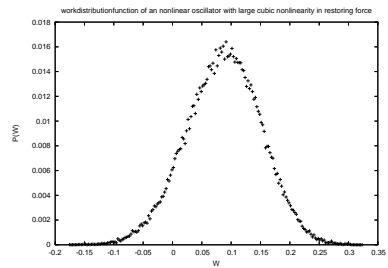}
\caption{Here force is applied for 8.4 sec. $M_0$ and $\tau$ are 100 and 840 respectively, in S.I. units. Here $\lambda=20$. The symmetrical nature is due to dominance of $\left<(W-\left<W\right>)^4\right>$ of the distribution here. Here $V(x)=(1/2)x^2+(20/4)x^4$.}
\end{figure}

\begin{figure}[b]
\centering
\includegraphics[scale=0.2,angle=270]{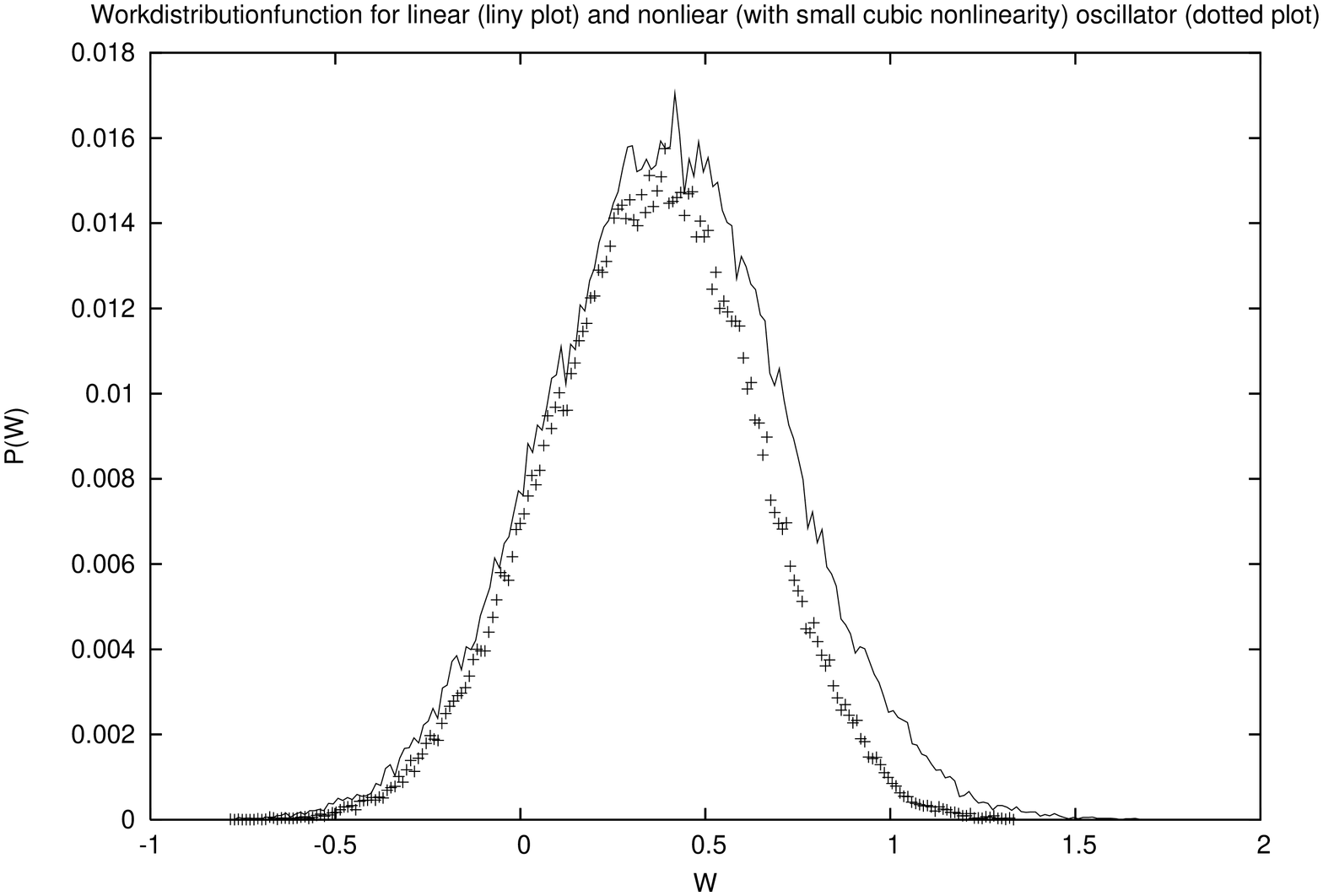}
\caption{For same values of $M_0$, $\tau$ and the time of application of $M(t)$ as before, but for $V(x)=(1/2)x^2$ and $V(x)=(1/2)x^2+(0.1/4)x^4$ the above liny and dotted plots are done respectively. The striking asymmetry in nonlinear case is clear here.}
\end{figure}



\begin{thebibliography}{99}
\bibitem{1}C.Jarzynski \prl {78} {2690} {1997}  
\bibitem{2}C.Jarzynski \pre {56} {5018} {1997}
\bibitem{3}C.Jarzynski \jsp {98} {77} {2000}
\bibitem{4}G.E.Crooks  \pre {60} {2721} {1999}
\bibitem{5}G.E.Crooks  \pre {61} {2361} {2000}
\bibitem{6}F.Douarche, S.Ciliberto, A.Patrosyan I.Rabbiosi \epl {70} {593} {2005}
\bibitem{7}F.Douarche, S.Ciliberto, A.Petrosyan \jsm {}{2005}{P09011}
\bibitem{8}Statistical Physics by L. D. Landau and E. M. Lifshitz Pergamon Press{1959}
\bibitem{9}F.Douarche, S.Joubaud, N.B.Garnier, A.Petrosyan and S.Ciliberto. \prl {97} {140603} {2006}
\bibitem{10}arXiv:condmat/{0612021}     
\end{thebibliography}
\end{document}